\documentclass[aps,prb,twocolumn,floatfix,showpacs]{revtex4}
\usepackage{amsmath,amsthm}  
\usepackage{epsf}
\usepackage{amssymb}  
\usepackage{graphicx}

\usepackage{bbm}
\usepackage{bm}
\newcommand{\ket}[1]{| #1 \rangle\rangle}
\newcommand{\bra}[1]{\langle\langle \tilde{#1} |}

\newcommand{\be}{\begin{eqnarray}}
\newcommand{\ee}{\end{eqnarray}}

\newcommand{\eq}[1]{Eq.~(\ref{#1})}

\begin{document}
\title{Waiting Times and  Noise in Single Particle Transport}
\author{Tobias Brandes}
\affiliation{  Institut f\"ur Theoretische Physik,
  Hardenbergstr. 36,
  TU Berlin,
  D-10623 Berlin,
  Germany}
\begin{abstract}
The waiting time distribution $w(\tau)$, i.e. the probability for a delay $\tau$ between two subsequent transition (`jumps') of particles, is a statistical tool in (quantum) transport. Using generalized Master equations for systems  coupled to external particle reservoirs, one can establish relations between  $w(\tau)$ and other statistical transport quantities such as the noise spectrum and the Full Counting Statistics. It turns out that $w(\tau)$ usually contains additional information on system parameters and properties such as quantum coherence, the number of internal states, or the entropy of the current channels that participate in transport. 
\end{abstract}
\maketitle

\section{Introduction}
Particle transport through a given open system $S$ can be regarded as a form of spectroscopy: by monitoring the flow of particles  between the system and external particle reservoirs (usually in a stationary state), one likes to extract as much information about $S$ as possible. Take $S$ as a `black box' and the  series of times $t_i^{\alpha}(i=1,2,...) $, where  individual particles leave or enter $S$ through reservoir $\alpha$, as the only available data. Just by counting (and perhaps labeling according to energy, spin etc...), one would  like to `reconstruct' $S$, thus defining an inverse problem where (some) transport data are known but the system Hamiltonian or Liouvillian is not.

Of course, part of this scenario is just a description of modern (quantum) transport, i.e. the determination of  full counting statistics (FCS)  or higher cumulants, current fluctuations, or noise spectra $S(\omega)$, rather than the determination of, say,  `just' the stationary current or a current-voltage characteristics. 

In this paper, the waiting time distribution $w(\tau)$ as a statistical tool for transport (described by generalized Master equations) is analyzed for systems with typically a finite number of discrete internal states (including possible quantum coherences), among which transitions occur due to the coupling to external reservoirs. $w(\tau)$ is a probability density for the delay time $\tau$ between two subsequent `jump' events (e.g. single electrons tunneling out of a quantum dot) and is well-known from quantum optics \cite{Caretal89}. 
In quantum transport, it has occasionally been used in the discussion of shot noise \cite{DHHW92}, counting statistics \cite{WEHM07,EL08} and in the description of transport through single, vibrating molecules \cite{KRO05,KOA06}. The purpose of this paper is to find out in how far the waiting time distribution contains information that is (partly) complementary to the one contained in, e.g., FCS and $S(\omega)$. For example, these latter can be derived from $w(\tau)$ but not vice-versa.

In particular, the waiting times contain spatially separate information on, e.g.,  tunnel barriers in quantum dots, and they depend  on the number of internal transitions, the quantum coherence, and the single or multiple `reset' character of the system, i.e. the number of system states that are directly involved in the process of particles entering or leaving the system. The simplest example of a single-reset system in quantum optics is resonance fluorescence in two-level atoms, where each emitted photon leaves the atom in its ground state. Corresponding examples in electron transport are quantum dots in the regime of strong Coulomb blockade, where only tunneling of one additional `transport' electron is possible. As a consequence, $w(\tau)$ then has to vanish for $\tau\to 0$ when measured at the same terminal.

Experimentally, various groups \cite{Gusetal06,Fujetal06} have achieved to measure time-series of single electron tunneling events, e.g. by monitoring transport through quantum dots using a nearby quantum point contact. One has to point out, however, that the formalism developed here always refers to counting the additional (or missing) particles in the reservoirs and not in the system $S$. Counting transitions within $S$ by direct interaction of a counting (measuring device) with $S$ usually would destroy quantum coherent features within $S$.

The paper is organized as follows: the next section introduces the general method on a more formal level which is introduced for  generalized Master equations. In the third section, various example applications  and  the entropy of currents as a possible way to distinguish between different classes of systems are discussed. Some remarks are left for the conclusions.

\section{Method}
There is a large number of cases where one can describe quantum transport in terms of a Markovian generalized Master equation for the reduced system density operator $\rho(t)$,
\be
\label{eofmotion}
\dot{\rho}(t) = \mathcal{L}\rho(t),\quad \mathcal{L}=\mathcal{L}_0+\mathcal{L}_1,\quad \mathcal{L}_1=\sum_{k=1}^M \mathcal{J}_k.
\ee
Here, {$\mathcal{L}_1$ describes $M$ different types of jump processes where single (quasi) particles tunnel into or out of the system.} 
Most of the applications below refer to single electron tunneling, but the formalism stays valid as long as one has a time evolution of the form \eq{eofmotion}.

The individual jump processes depend on the specific terminal (e.g., left and right), energy level, system state before and after the jump etc., but at this stage one makes no further assumption other than that the jump operators act on density matrices $\rho$ as 
\be
\label{Jkdef}
\mathcal{J}_k\rho = c_k[\rho] \rho_k,
\ee
where $\rho_k$ is a density matrix and $c_k[\rho]$ is a scalar that is typically given in terms of positive transition rates.

The splitting $\mathcal{L}=\mathcal{L}_0+\mathcal{L}_1$ is not unique and depends very much on the kind of information one wishes to obtain from a transport measurement. A large $M$ means that one can access a large number of jump processes of different kind (e.g., energy resolved), but this is typically very difficult to monitor experimentally. Also numerically, there is a trade-off between making $\mathcal{L}_0$ as diagonal as possible by having a large $M$, leading to large but simply structured  matrices for the waiting times and a corresponding large inversion problem for the noise spectrum, and having small $M$, leading to small waiting time matrices with, however, a relatively complicated structure (see below).

In the interaction picture  with respect to $\mathcal{L}_0$ and  $S_t\equiv  e^{\mathcal{L}_0 t}$,  a formal solution (`unraveling') of the Master equation \eq{eofmotion} reads
\be
\rho(t) = \sum_{n=0}^{\infty}   \sum_{l_1=1,...,l_n=1}^M   \int_0^{t} dt_n ...\int_0^{t_{2}}dt_{1} 
\rho^t_{l_n...l_1}(\{t_i\}),
%
\ee  
where the
\be
\label{rho_cond}
\rho^t_{l_n...l_1}(\{t_i\}) \equiv S_{{t}-t_n} \mathcal{J}_{l_n}  S_{t_n-t_{n-1}} \mathcal{J}_{l_{n-1}}  ... \mathcal{J}_{l_1}  S_{t_1} \rho_{\rm in}
\ee
are (unnormalized) conditioned density operators \cite{Carmichael} that  describe the non-unitary time-evolution of an initial density operator $\rho_{\rm in}$, interrupted by $n$ quantum jumps of type $l_i$ at times $t_i$ with $i=1,...,n$. 

In the following, the existence of a unique stationary solution $\rho_0$ of \eq{eofmotion} is always assumed, i.e. decomposable Liouvillians  \cite{vanKampen} with a block structure referring to the uninteresting case of two (or more) de-coupled systems are excluded.  For $n=2$ subsequent jumps of type $k$ and $l$ with $t_1=0$ and $t_2=\tau=t$, one now defines  the normalized conditioned density operator $\rho^c_{kl}$ with  $\rho_{\rm in}=\rho_0$ in \eq{rho_cond},
\be
\rho^c_{kl}(\tau) \equiv \frac{\mathcal{J}_k   e^{\mathcal{L}_0 \tau}\mathcal{J}_l  \rho_0}{w_{kl}(\tau)\mbox{\rm Tr} \mathcal{J}_l \rho_0 }.
\ee
The normalization factor in the denominator defines a matrix $\mathbf{W}(\tau)$ of waiting time distributions
\be
\label{wdefinition}
w_{kl}(\tau)\equiv \left(\mathbf{W}(\tau)\right)_{kl}\equiv
\frac{\mbox{\rm Tr} \mathcal{J}_k   e^{\mathcal{L}_0 \tau}\mathcal{J}_l  \rho_0}{{I}_l  },
\ee
where 
\be
\label{current_def}
I_l\equiv \mbox{\rm Tr} \mathcal{J}_l  \rho_0
\ee
defines a stationary current due to jump processes of type $l$. Here and in the following, all currents have physical dimension $1/$time.

\eq{eofmotion} defines a stochastic process that is described by a linear system of coupled first order differential equations. For practical calculations, it is therefore convenient to  represent the density operator $\rho$ as a column vector with real entries,  for example as $\rho=(\rho_{11},\rho_{22},...,\rho_{NN},\Im \rho_{12},\Re \rho_{12},...,
\Re \rho_{NN-1})^T$. Correspondingly, the super-operators $\mathcal{J}_k$, $\mathcal{L}$, $\mathcal{L}_0$ become real matrices. The next step is to introduce a convenient Dirac-like notation \cite{FNJ04} where kets $\rho \leftrightarrow \ket{\rho}$ denote normalized density operators. The stationary state $\rho_0$ with $\mathcal{L}\rho_0=0$ is denoted as ket $\ket{0}$.
The $\bra{0}$ is the row vector $(1,1,...1,0,0...0)$ such that in this vector representation of density operators $\rho$, the trace operation on a column vector $A \rho$ (where $A$ is an arbitrary super-operator) becomes the scalar product $\mbox{\rm Tr} A \rho= \bra{0} A \rho\rangle\rangle$. In particular, one has $ \bra{0}{\rho}\rangle\rangle=1$ for density operators $\rho$.
Finally, the bras $\bra{k}$ are real row-vectors defined via 
the action of the jump operators $\mathcal{J}_k$, \eq{Jkdef}, {when writing their matrices as the dyadic product}
\be
\mathcal{J}_k \equiv \ket{k}\bra{k}.
\ee
Note that bras (row vectors) $\bra{k}$ and kets (column vectors) $\ket{k}$ are independent vectors and are not dual to each other. They are introduced here as a convenient notation for the calculations to follow. They  have positive entries as the transitions rates for the quantum jumps are positive. 

As the diagonal elements of the Liouvillian $\mathcal{L}$ do not describe jump processes but the conservation of probability, the matrices $\mathcal{J}_k$ have no diagonal elements, and consequently $\bra{k} k\rangle\rangle=0$ for $k\ne 0$ and $\mathcal{J}_k^2=0$.
By furthermore taking the trace in \eq{Jkdef}, one has $c_k[\rho]= \mbox{\rm Tr} \mathcal{J}_k\rho = \bra{k} \rho\rangle \rangle$ and also in \eq{wdefinition},
\be
w_{kl}(\tau)&\equiv& \frac{\bra{k}   e^{\mathcal{L}_0 \tau}  \ket{l}\bra{l} 0 \rangle\rangle}{  \bra{l} 0\rangle\rangle } =
 \bra{k}   e^{\mathcal{L}_0 \tau}  \ket{l},
\ee
where the stationary currents $I_l= \bra{l} 0\rangle\rangle $ due to jump processes of type $l$ cancel, cf. the definition \eq{wdefinition}. As a consequence, the calculation of the waiting times $w_{kl}(\tau)$ does not require the knowledge of the stationary state, $\rho_0$. Note that the $w_{kl}(\tau)$ are manifestly real quantities. {With $\bra{k}$ and $ \ket{l}$  having positive entries}, the $w_{kl}(\tau)$ are also positive, provided  that  the time-evolution operation $e^{\mathcal{L}_0 \tau}$ does not lead to non-positive definite density operators - a condition which is physically plausible and which is fulfilled in all the examples discussed below.

One now defines a waiting time super-operator $\mathcal{W}_l(z)$ as the Laplace transform of $e^{\mathcal{L}_0 \tau}\mathcal{J}_l $,
\be
\mathcal{W}_l(z)\equiv (z-\mathcal{L}_0)^{-1} \mathcal{J}_l ,
\ee
by which the  Laplace transform $\hat{w}_{kl}(z)$ of the  waiting time distributions can be conveniently written as an `expectation value', 
\be
\label{w_Laplace}
\hat{w}_{kl}(z) &\equiv &\int_0^{\infty} dte^{-zt} w_{kl}(t) = \frac{ \mbox{\rm Tr} \mathcal{J}_k\mathcal{W}_l(z)\rho_0}{\mbox{\rm Tr} \mathcal{J}_l \rho_0}\nonumber\\
&=& \bra{k}  (z-\mathcal{L}_0)^{-1} \ket{l}.
\ee

For any initial jump of type $l$, the waiting time distribution  $w_{kl}(t)$ has to give unity when summed over all subsequent jumps of type $k$ and integrated over all times $\tau$. This normalization indeed follows from 
\be\label{normalisation}
& &\int_0^{\infty} dt \sum_k w_{kl}(t)= \sum_k \hat{w}_{kl}(0) = \\
&=& - \frac{\bra{0} (\mathcal{L}-\mathcal{L}_0)\mathcal{L}_0^{-1} \mathcal{J}_l \ket{0}}{I_l} 
= \frac{\bra{0}\mathcal{J}_l   \ket{0}}{I_l} = 1,\nonumber
\ee
where one exploits the fact that $\bra{0}$ is a left eigenvector with eigenvalue zero of the total Liouvillian,
$\bra{0}\mathcal{L}=0$, which reflects conservation of probability.

\subsection{Relation to the Noise Spectrum}
(Quantum) fluctuations of the system can be visualised in fluctuations of the current and play a central role in analysing the internal system dynamics via transport spectroscopy \cite{Bue92,BB00,Naz03,Bra05}. The spectrum 
\be
S_{kl}(\omega) \equiv \frac{1}{2}\int_{-\infty}^\infty dt e^{i\omega t} \langle \{ \delta I_k(t) ,  \delta I_l(0) \} \rangle
\ee
defines fluctuations of currents, where $\delta I_k(t)$ denotes the deviation of current $I_k$ from its average in the stationary state, \eq{current_def}.  It 
is derived via the MacDonald formula \cite{EG02,FNJ04,CBB04,LAB07},
\be
\label{novotny_formula}
S_{kl}(\omega) &=& \delta_{kl}I_l \\
&-&  \frac{1}{2} \left\{\sum_\pm\mbox{\rm Tr}\mathcal{J}_k \frac{1}{\pm i\omega + \mathcal{L}}  \mathcal{J}_l  \rho_0 +(k\leftrightarrow l)\right\},\nonumber
\ee
which for numerical convenience can also be expressed in terms of a resolvent operator \cite{FNJ04} 
$ R(\omega) \equiv i{\ket{0}\bra{0}}/{\omega} + Q[{i\omega + \mathcal{L}}]^{-1}Q$
with $Q= 1- \ket{0}\bra{0}$ and where the singular contribution from the stationary solution at $\omega=0$ has been projected out.

The link with the waiting time distributions is now established via the operator identity 
\be
\label{opindent}
 \left(z-\mathcal{L}\right)^{-1} \mathcal{J}_l &=&  \left[ \left( z-\mathcal{L}_0 \right) \left( 1- \left( z-\mathcal{L}_0 \right)^{-1}\mathcal{L}_1 \right)  \right]^{-1} \mathcal{J}_l \nonumber\\ 
&=& \left[1- \sum_{m=1}^M\mathcal{W}_m(z)  \right]^{-1} \mathcal{W}_l(z).
\ee
The expression $\mathcal{J}_k\left(z-\mathcal{L}\right)^{-1} \mathcal{J}_l$
is then formally expanded into a geometric series, the first term of which is given by
\be
\bra{0} \mathcal{J}_k \mathcal{W}_l(z) \ket{0} &=& 
\bra{k} (z- \mathcal{L}_0)^{-1} \ket{l}\bra{l} 0\rangle\rangle \nonumber\\
&=& \hat{w}_{kl}(z)  {I_l} \equiv  I_l \left[\mathbf{W}(z)\right]_{kl},
\ee
where $\mathbf{W}(z)$ is the matrix of the Laplace transformed $\hat{w}_{kl}(z)$.
The $n+1$-th term of the geometric series contains correspondingly
\be
\bra{k} \mathcal{W}_{m_1}(z)... \mathcal{W}_{m_n}(z) \mathcal{W}_l(z)  \ket{0} =\\
= 
\hat{w}_{km_1}(z) \hat{w}_{m_1m_2}(z)  ... \hat{w}_{m_n l}(z){I_{l}}
\ee
which upon summation simply yields matrix products, i.e. the $n+1$-th power of the matrix $\mathbf{W}$,
and thus
\be\label{sumgeo}
& & \bra{0} \mathcal{J}_k \left(z-\mathcal{L}\right)^{-1} \mathcal{J}_l \ket{0} = \\
&=& \bra{k}  \sum_{n=0}^{\infty} \left(\sum_{m=1}^M\mathcal{W}_m(z) \right)^n \mathcal{W}_l(z)\ket{0}\nonumber\\
&=& \sum_{n=0}^{\infty} I_l \left[\mathbf{W}^{n+1}(z)\right]_{kl} 
=  I_l \left[\left(1-\mathbf{W}(z)\right)^{-1}\mathbf{W}(z)\right]_{kl}\nonumber.
\ee
Up to here, this is a formal result as the sum \eq{sumgeo} does not need to converge (the expression actually has to be singular for $z=0$ due to the normalization of the waiting times). For purely imaginary $z=i\omega$ with real $\omega\ne 0$, however, for positive ${w}_{kl}(t)\ge 0$ and due to the normalization \eq{normalisation}, the matrix norm
\be
\| \mathbf{W}(i\omega) \| &\equiv& \max_l
\sum_k \left| \hat{w}_{kl}(i\omega)\right| \nonumber\\
&\le&   \max_l \sum_k\int_0^{\infty}dt {w}_{kl}(t)=1,
\ee
and with $\| \mathbf{W}^n(z) \|\le\| \mathbf{W}(z)\|^n $ it follows that the series converges.  

Purely imaginary $z=i\omega$ is just what is needed in Eq. (\ref{novotny_formula}), and one obtains 
\be
\label{novotny_formula_w}
& & {S_{kl}(\omega)}= \delta_{kl} I_l\\
&+& \frac{1}{2} \left\{ \sum_\pm  \left[\left(1- \mathbf{W}(\pm i\omega)  \right)^{-1} \mathbf{W}(\pm i\omega) \right]_{kl} I_l +(k\leftrightarrow l)\right\},\nonumber
\ee
which expresses the noise spectrum in terms of the waiting time distributions.

\subsection{Relaxation Currents}
After a quantum jump of type $l$, the system relaxes from the reset state $\rho_l\equiv\ket{l}$ into the stationary state. Such a relaxation potentially involves all jump processes of type $k$, and consequently relaxation currents should be defined as  
\be
\label{relaxdef}
I_{kl}^{\rm relax}(t) &\equiv& \mbox{\rm Tr}\mathcal{J}_k e^{\mathcal{L}t } \rho_l = \bra{k} e^{\mathcal{L}t }  \ket{l}.
\ee
Upon Laplace-transformation, one finds
\be
\mbox{\rm Tr} \mathcal{J}_k \left(z-\mathcal{L}\right)^{-1} \mathcal{J}_l\rho_0 &=& 
\bra{k} \left(z-\mathcal{L}\right)^{-1} \ket{l} \bra{l}\ket{0}  \nonumber\\ 
&=&  \hat{I}_{kl}^{\rm relax}(z) I_l,\quad z\ne 0,
\ee
where $\hat{I}_{kl}^{\rm relax}(z)$ is the Laplace transform of $\langle I_{kl}^{\rm relax} \rangle(t)$, and {thus by comparison with \eq{novotny_formula}},
\be
\label{noise_bycomparison}
S_{kl}(\omega) &=& \frac{1}{2}{I_l}   \left\{\delta_{kl} + \sum_\pm \hat{I}_{kl}^{\rm relax}(\pm i\omega) \right\} +(k\leftrightarrow l).
\ee
The fluctuations of the stationary currents are thus determined by the relaxation currents from the reset states $\rho_l$. 
The relaxation currents are in turn determined by the waiting time distribution,
\be
\label{Iwrelation}
 \hat{I}_{kl}^{\rm relax}(z) =  \left[\left(1- \mathbf{W}(z)  \right)^{-1} \mathbf{W}(z) \right]_{kl}.
\ee
In the time domain, the relaxation currents are obtained from the waiting times by re-expanding \eq{Iwrelation} as a series of convolution integrals,
\be
\label{Irelaxexpansion}
{I}_{kl}^{\rm relax}(t)&=&\nonumber\\
 w_{kl}(t) &+& \int_0^tdt_1\left[  \mathbf{W}(t-t_1) \mathbf{W}(t_1)\right]_{kl} +... 
\ee

Eqs.  (\ref{noise_bycomparison}), (\ref{Iwrelation}) also illustrate a numerical trade-off when calculating waiting time distributions and noise spectra for different unravelings of $\mathcal{L}$, \eq{eofmotion}:
splitting a large number $M$ of jump operators off the total Liouvillian $\mathcal{L}$ has the benefit of potentially simple  Liouvillians $\mathcal{L}_0$, e.g. close to lower-triangular form. As a downside, this leads to large waiting time matrices $ \mathbf{W}(z)$ and therefore to a potentially large matrix inversion problem for the determination of the $\hat{I}_{kl}^{\rm relax}(z)$ when calculating the noise spectrum.

\subsection{Single and Multiple Reset Systems}
The individual jump operators $\mathcal{J}_k$, \eq{Jkdef}, refer to a unique density matrix $\ket{k}$ after the jump of type $k$. One can now further classify jump processes by introducing  class labels $\alpha$ by slightly extending the notation,  writing  double indices $k\alpha$ in 
\be
\mathcal{L}=\mathcal{L}_0 + \mathcal{L}_1,\quad \mathcal{L}_1=\sum_{k\alpha} \mathcal{J}_{k\alpha}
\ee
with  $ \mathcal{J}_{k\alpha} \equiv \ket{k\alpha}\bra{k\alpha} $, cf. \eq{Jkdef}. 
In the examples below, $\alpha=L/R$ labels  two  (left and right) particle reservoirs (leads) attached to the system. The index $k$ then further specifies the type of jump process for reservoir $\alpha$. Depending on the physical system described by $\mathcal{L}$, $k$ can label the jump process according to energy, spin, or some other  degrees of freedom that are coupled to the transport process, e.g. phonons. 

A simple counting without energy resolution of, e.g., the emitted electrons, is the situation which - in contrast to quantum optics - is typical of quantum transport experiments. One then is interested in  sums of jump operators $ \mathcal{J}_{k\alpha} $ only, i.e.  super-operators  $\mathcal{J}_{\alpha}$ and currents $I_\alpha$ 
\be\label{Jalphadef}
\mathcal{J}_\alpha \equiv \sum_{k} \mathcal{J}_{k\alpha},\quad I_\alpha \equiv \mbox{\rm Tr} \mathcal{J}_{\alpha} \rho_0.
\ee
Individual jump processes within the class $\alpha$  typically leave the system in different states $\ket{k\alpha}$. Some systems have a unique `reset' state $\ket{\alpha}$, as for example the empty state of a quantum dot in the strong Coulomb blockade regime with only one or zero electrons. This is  in contrast to other cases where for example the system particle number fluctuates by more than one in the stationary state. In general, this distinction leads to the definition of single and multiple `reset' systems: one defines a single reset system as a system 
where all jump operators $\mathcal{J}_\alpha=\sum_{k} \mathcal{J}_{k\alpha}$ have a reset property analogous to \eq{Jkdef}, i.e.
\be
\mathcal{J}_\alpha\rho &=& c_\alpha[\rho] \rho_\alpha.
\ee
This condition is fulfilled if the state $\ket{k\alpha}= \ket{\alpha}$ independent of $k$, i.e.
\be
\label{sepdef}
\mathcal{J}_\alpha = \sum_{k} \ket{k\alpha}\bra{k\alpha} =\ket{\alpha}\sum_{k}\bra{k\alpha}\equiv \ket{\alpha}\bra{\alpha}.
\ee

In general, a sum over jump operators $\mathcal{J}_k$ can {\em not} be written in `separable' form \eq{sepdef}, but the bras and kets are `entangled', and  the corresponding system is called a `multiple reset' system.

In analogy with the definition \eq{w_Laplace} of the waiting time distribution, the quantity
\be
\label{wreducedef}
 \hat{w}^{(r)}_{\alpha\beta}(z)\equiv  \frac{\mbox{\rm Tr} \mathcal{J}_\alpha \mathcal{W}_\beta(z)\rho_0}{  \mbox{\rm Tr} \mathcal{J}_\beta\rho_0   }
\ee
defines a (reduced) waiting time distribution which involves the $\mathcal{J}_\alpha$, \eq{Jalphadef}, only. Accordingly, one defines
 the reduced  noise spectrum matrix of the currents $I_\alpha$ and $I_\beta$, 
\be
\label{Sreduced}
& &S^{(r)}_{\alpha\beta}(\omega) \equiv   \delta_{\alpha\beta}I_\beta \\
&-&  \frac{1}{2} \left\{ \sum_\pm\mbox{\rm Tr}\mathcal{J}_\alpha \frac{1}{\pm i\omega + \mathcal{L}}  \mathcal{J}_\beta  \rho_0
 + (\alpha \leftrightarrow \beta) \right\} \nonumber\\
&=&\frac{1}{2} I_{\beta}\left\{ \delta_{\alpha\beta} + \sum_\pm \hat{I}_{\alpha\beta}^{\rm relax}(\pm i\omega) \right\}
+ (\alpha \leftrightarrow \beta)  \nonumber
\ee
with  $S^{(r)}_{\alpha\beta}(\omega)= \sum_{k l}S_{k\alpha,l\beta}(\omega) $ and 
the {\em reduced relaxation currents} 
\be\label{Irelaxreduce}
 \hat{I}_{\alpha\beta}^{\rm relax}(z) &\equiv& 
\frac{\sum_{kl}   \left[\left(1- \mathbf{W}(z)  \right)^{-1} \mathbf{W}(z) \right]_{k\alpha,l\beta} I_{l\beta}}
{\sum_{l} I_{l\beta}}.
\ee
Analysing this definition, one recognizes a profound difference between single and multiple reset systems in their respective connections between noise and waiting time distribution:
in single reset systems, due to the uniqueness $\ket{l\beta}=\ket{\beta}$  one has
\be
\hat{w}^{(r)}_{\alpha\beta}(z)&=& \frac{ \sum_{k l }\bra{k\alpha}(z-\mathcal{L}_0)^{-1}\ket{l\beta} I_{l\beta}}{ \sum_{l}I_{l\beta}} \nonumber\\
&=&  \bra{\alpha}(z-\mathcal{L}_0)^{-1}\ket{\beta}.
\ee
The reduced relaxation currents then simplify drastically: re-expanding the geometric series in \eq{Irelaxreduce}, 
\be
& & \hat{I}_{\alpha\beta}^{\rm relax}(z) =  \sum_{k} \bra{k\alpha} (z-\mathcal{L}_0)^{-1}\ket{\beta} + \nonumber\\
&+& \sum_{k}\sum_{l'\alpha' }   \bra{k\alpha} (z-\mathcal{L}_0)^{-1}\ket{\alpha'}  \bra{l'\alpha'} (z-\mathcal{L}_0)^{-1}\ket{\beta} +...\nonumber\\
&=&  \hat{w}^{(r)}_{\alpha\beta}(z) + \sum_{\alpha'}  \hat{w}^{(r)}_{\alpha\alpha'}(z)  \hat{w}^{(r)}_{\alpha'\beta}(z) +...\
\ee
where the sum over $l$ cancels the currents $\sum_l I_{l\beta}$, and therefore
\be
\label{Isimplyrelax}
 \hat{I}_{\alpha\beta}^{\rm relax}(z) &=&  \left[\left(1- \mathbf{W}^{(r)}(z)  \right)^{-1} \mathbf{W}^{(r)}(z) \right]_{\alpha\beta},
\ee
with the matrix $\mathbf{W}^{(r)}(z)$ of the {\em reduced} waiting time distribution, \eq{wreducedef}.
Therefore in single reset systems, the reduced noise spectrum matrix $S^{(r)}_{\alpha\beta}(\omega)$ and the reduced waiting time distribution $\hat{w}^{(r)}_{\alpha\beta}(\omega)$ are connected via a matrix relation, cf. \eq{Isimplyrelax} and \eq{Sreduced}, involving the matrix $\mathbf{W}^{(r)}(z)$ which in general has a smaller size than the full waiting time matrix $\mathbf{W}(z)$.

In the non-separable case, i.e. where the super-operator for the total current through the system can not be written as in \eq{sepdef}, this is no longer the case, and one has to use the full matrix equation \eq{novotny_formula_w} in order to relate both quantities.

\subsection{High-frequency and short time expansions}\label{highfreq}
The noise formula, \eq{novotny_formula_w}, can formally be expanded in powers of $\hat{w}_{kl}(z)$,
\be
\label{expansionlargew}
{S_{kl}(\omega)} &=& \frac{1}{2} \left\{\delta_{kl} + \sum_\pm  \hat{w}_{kl} (\pm i\omega) + O( \hat{w}^2) \right\} I_l \nonumber\\
&+& (k\leftrightarrow l),
\ee
where the terms involving  first and higher powers of $\hat{w}_{kl}(z)$ describe the deviation from the Poissonian limit $S_{kl}(\omega) = \delta_{kl}I_l$. In actual fact, this limit is reached for $\omega\to \infty$ since $\lim_{z\to\infty} \hat{w}_{kl}(z) =0$,
and one therefore expects \eq{expansionlargew} to be a high-frequency expansion. 
In particular, one has for $z\to\infty$,
\be
 \hat{I}_{kl}^{\rm relax}(z) \to \hat{w}_{kl}(z),\quad z\to \infty.
\ee
The large $z$ limit corresponds to small waiting times $\tau$, and for small $\tau$ the system has not enough time to resolve all its possible states after a quantum jump. At small times $t$, relaxation currents and waiting times thus coincide,
\be
 {I}_{kl}^{\rm relax}(\tau) \to {w}_{kl}(\tau),\quad \tau\to 0,
\ee
{which also follows from \eq{Irelaxexpansion}.} 

One obtains the short time expansion of ${w}_{kl}(\tau)$ from an expansion of $\hat{w}_{kl}(z)$ into a Laurent series. For example, when 
\be
\label{shorttimeform}
{w}_{kl}(\tau) \sim c_{kl}\tau^{n-1},\quad \tau\to 0
\ee
with $n>0$, one has $\hat{w}_{kl}(z\to\infty)\sim c_{kl}\Gamma(n) z^{-n}$,
where $\Gamma(.)$ is the Gamma function. 
Alternatively, the exponent $n$ is directly obtained  by using the expansion
\be
\label{shorttimedirect}
{w}_{kl}(\tau) &\equiv& \bra{k} e^{\mathcal{L}_0\tau} \ket{l}=\nonumber\\
&=&   \bra{k}   1+ \mathcal{L}_0 \tau + \frac{\tau^2}{2}\mathcal{L}_0^2+...   \ket{l}.
\ee
In the examples below, one finds that the asymptotic behaviour of ${w}_{kl}(\tau)$ at small $\tau$, i.e. the exponent $n$, 
depends on the dynamics within the system and thus contains potentially useful information.

Finally, the high frequency expansion of the  reduced noise spectrum, \eq{Sreduced}, reads 
\be\label{highfreduced}
S^{(r)}_{\alpha\beta}(\omega) 
&=& \frac{1}{2}I_\beta  \left\{  \delta_{\alpha\beta}  + \sum_\pm  \hat{w}^{(r)}_{\alpha\beta}(\pm i\omega)  +...\right\} \nonumber\\
&+& (\alpha \leftrightarrow \beta),
\ee
which means that at large frequencies, the correction to $S^{(r)}_{\alpha\beta}(\omega)$  to the Poissonian limit is directly given by the reduced waiting time distribution regardless of whether or not the system is of single reset type.

\subsection{Relation to Full Counting Statistics (FCS)}
For any given unraveling $\mathcal{L}=\mathcal{L}_0+\mathcal{L}_1$, \eq{eofmotion}, one expects the waiting time distribution to be related to  the Full Counting Statistics (FCS), i.e. the probability $p(n,t)$ of $n$ quantum jumps in a time interval $[0,t]$. For a single reset system and a single jump operator  $\mathcal{L}_1=\mathcal{L}\equiv \ket{1}\bra{1}$,  the generating function $G(\chi,t)$ associated with $p(n,t)$ is obtained from the counting-variable ($\chi$)-dependent propagator \cite{Cook81,Len82,BN03,Emetal07},
\be
G(\chi,t)\equiv \sum_{n=0}^\infty e^{i n\chi}p(n,t)=\mbox{\rm Tr} e^{(\mathcal{L}_0+ e^{i \chi}\mathcal{J})t}\rho_0,
\ee
where $\rho_0$ again denotes the stationary state. 
The Laplace transform $\hat{G}(\chi,z)$ of the generating function therefore is 
\be
\label{FCS_wait1}
\hat{G}(\chi,z) &=&  \bra{0} (z- \mathcal{L}_0  -  e^{i \chi}\mathcal{J}    )^{-1}\ket{0}\nonumber\\
&=& \sum_{n=0}^\infty \bra{0} \left[(z- \mathcal{L}_0)^{-1}  e^{i \chi}\mathcal{J}\right]^{n}  (z- \mathcal{L}_0)^{-1}  \ket{0}\nonumber\\
&=& \hat{w}_{00}(z) + \frac{ \hat{w}_{01}(z) \hat{w}_{10}(z)}{e^{-i \chi} - \hat{w}_{11}(z) },
\ee
with the definition
\be
\hat{w}_{ij}(z)&\equiv& \bra{i} (z- \mathcal{L}_0)^{-1} \ket{j},\quad i,j,=0,1.
\ee
The long-time behaviour of $p(n,t)$ is then obtained from the $\chi$-dependent pole of \eq{FCS_wait1}, i.e. the  solution $z_0$ of
\be
\label{FCS_single}
e^{-i \chi} - \hat{w}_{11}(z_0) =0
\ee
with $z_0(\chi=0)=0$ (note that the normalization of $w_{11}$ is  $\hat{w}_{11}(0)=1$), and derivatives of the function $z_0(\chi)$ then yield all the cumulants of $p(n,t\to \infty)$ \cite{BN03,Emetal07}.

In the multiple reset case or in general for  more than one jump operator, one has $\mathcal{L}_1=\sum_{k=1}^M \mathcal{J}_k$, \eq{eofmotion}, and correspondingly $M$ counting variables $\chi_k$, but the derivation of $G(\{\chi_k\},z)$ is analogous to the above one. The $\chi$-dependent propagator
is $\mathcal{L}(\{\chi_k\}) =  \mathcal{L}_0+ \sum_{k=1}^M \mathcal{J}_k e^{i \chi_k}$ and the generating function is 
\be
G(\{\chi_k\},z) &=& \hat{w}_{00}(z) + \mathbf{u}^T \left[e^{-i\chi} - \mathbf{W}(z) \right]^{-1}  \mathbf{v},
\ee
where $e^{-i\chi}$ is the diagonal matrix of the $e^{-i\chi_k}$ and the vectors $\mathbf{u}$ and $\mathbf{v}$ have components
\be
\left[\mathbf{u}\right]_k &=& \bra{0} (z- \mathcal{L}_0)^{-1} \ket{k} \nonumber\\
\left[\mathbf{v}\right]_k &=& \bra{k} (z- \mathcal{L}_0)^{-1} \ket{0},\quad k=1,...,M.
\ee
The condition 
\be
\label{FCS_multiple}
\det\left[e^{-i\chi} - \mathbf{W}(z) \right]=0
\ee
then defines a polynomial in $z$, of which the zero $z_0(\{\chi_k\})$ with $z_0(0)$ determines the FCS.

\section{Information Contained in the Waiting Time Distribution, Examples}\label{example_section}

From the considerations so far, two questions arise: first, what additional information (as compared with the noise spectrum $S_{kl}(\omega)$) does the waiting time distribution ${w}_{kl}(\tau)$ contain at all. Clearly, the noise spectrum  is an even function of $\omega$ whereas $\hat{w}(\omega)$ in general is not. In the examples below it is shown that indeed $\hat{w}(\omega)$ can reveal information that $S(\omega)$ does not contain.

 Second, the distinction between various types of jump processes gave rise to a matrix structure of both, the noise spectrum and the waiting time distribution. When only certain combinations of jump operators are accessible experimentally (e.g., when electron tunnel is not resolved as a function of energy), the difference between single and multiple reset systems becomes important and noise and waiting time can contain complementary information.

Before  discussing specific examples,  two further useful concepts are introduced -  first, a rule that relates various waiting times for different splittings of $\mathcal{L}$, \eq{eofmotion}, and second, the entropy of a stationary current distribution.

\subsection{Waiting Times for Two Different Splittings}\label{sectionsplitting}
As mentioned above, the splitting of the Liouvillian $\mathcal{L}=\mathcal{L}_0+\mathcal{L}_1$, \eq{eofmotion}, is not unique. The set of jump operators  $\mathcal{J}_k$ that are included in $\mathcal{L}_1=\sum_{k=1}^M \mathcal{J}_k $ defines the jump processes that one wishes to monitor.

Consider two splittings involving only one and two jump operators,
\be
\mathcal{L}=\mathcal{L}_0 + \mathcal{J}_1 +\mathcal{J}_2=\tilde{\mathcal{L}}_0 + \mathcal{J}_1.
\ee
The corresponding waiting time distributions are denoted as $w_{ij}(\tau)$ and $\tilde{w}(\tau)$. Using $\mathcal{J}_i\equiv \ket{i}\bra{i}$, the relation between these is found via
\be
\label{twosplit}
\hat{\tilde{w}}(z) &=& \bra{1} (z- \tilde{\mathcal{L}}_0)^{-1} \ket{1}\nonumber\\
&=& \bra{1} \left[1-(z- \mathcal{L}_0)^{-1} \ket{2}\bra{2} \right ]^{-1}(z- \mathcal{L}_0)^{-1} \ket{1}\nonumber\\
&=& \hat{w}_{11}(z) +\frac{\hat{w}_{12}(z)\hat{w}_{21}(z)}{1-\hat{w}_{22}(z)}.
\ee
Two subsequent type-1 jumps are separated by zero, one, two,... or any number of type-2 jumps. Therefore, the waiting time distribution $\tilde{w}(\tau)$ for type-1 jumps has to be given by the sum over all intermediate jump processes, which by re-expanding the geometric series is \eq{twosplit} in the time domain,
 \be
& &{\tilde{w}}(\tau)= 
{w}_{11}(\tau) + \int_0^\tau dt_1 \hat{w}_{12}(t-t_1) \hat{w}_{21}(t_1)\\
&+& \int_0^\tau dt_1 \int_0^{t_1} dt_2 \hat{w}_{12}(t-t_1)  \hat{w}_{22}(t_1-t_2)     \hat{w}_{21}(t_2)+...\nonumber
\ee
Relations similar to \eq{twosplit} can be derived when more than two jump operators are involved.

\subsection{Entropy of Current Distribution}
The currents $I_{\alpha}$ usually split into individual contributions $I_{k\alpha}\equiv \mbox{\rm Tr}\mathcal{J}_{k\alpha} \rho_0$, for example according to energy or spin. 
A useful concept to quantify the distribution of  the $I_{k\alpha}$ is the  entropy $E_\alpha$ of the individual currents $I_{k\alpha}$,
\be\label{current_entropy}
E_\alpha \equiv -\sum_k \frac{I_{k\alpha}}{I_\alpha} \log \frac{I_{k\alpha}}{I_\alpha}.
\ee
This definition is quite in analogy with the usual entropy  $\mathcal{S}[p_k]\equiv -\sum_k p_k \log p_k$ of a discrete distribution $p_k$.
These entropies, however, depend on the choice of the $I_{k\alpha}$ in the original unraveling. 

In a multiple reset system, a current $I_{\alpha}$ consists of at least two individual currents $I_{k\alpha}$, corresponding to the (at least two) reset states $\ket{k\alpha}$ whence the current entropy of a multiple reset system is non-zero. In contrast, in single reset systems there is always a choice  $I_{\alpha}=\ket{\alpha}\bra{\alpha}$, cf. \eq{sepdef}, with a single $\ket{\alpha}$ for which the entropies $E_\alpha$ are zero.

In the following, the above concepts are discussed for various examples.

\subsection{Transitions in a Ring}
Consider a ring with transitions between $N+1$ states, $0\to 1\to 2\to...\to N \to 0$ at rates $\Gamma_i$, $i=0,1,...,N$. The corresponding Liouvillian is
\be
\label{ring}
\mathcal{L} =\left(
  \begin{array}{ccccc}
-\Gamma_0 & 0 & ... & 0 &  \Gamma_N \\ 
 \Gamma_0 & -\Gamma_1 & 0 &... & 0\\
 0        & \Gamma_1 & -\Gamma_2 &...& 0\\
 ... &  ...  & ... & ... & ...\\  
 0  & 0 & ...       & \Gamma_{N-1} & -\Gamma_N\\
\end{array}\right).
\ee
For $N=1$ this is, for example, equivalent to transport of single electrons through a single level quantum dot (see below).

Introducing the single jump operator $\mathcal{J}\equiv\ket{1}\bra{1}$ with $\bra{1}\equiv(0,...,\Gamma_N)$ and $\ket{1}\equiv(1,0,...,0)^T$, and a splitting
$\mathcal{L}=\mathcal{L}_0+\mathcal{J}$ corresponding to measuring the jumps $N\to 0$ only, by inverting $z-\mathcal{L}_0$ one finds the waiting time distribution 
\be
\label{wz_chain}
 \hat{w}(z) &=& \bra{1} (z-\mathcal{L}_0)^{-1}\ket{1} = \Gamma_N \left[(z-\mathcal{L}_0)^{-1}\right]_{N0}\nonumber\\
            &=& \frac{\Gamma_0}{z+\Gamma_0} \frac{\Gamma_1}{z+\Gamma_1}...\frac{\Gamma_N}{z+\Gamma_N},
\ee
where  the matrix element of the inverse matrix is a ratio of two determinants. Therefore, the waiting time distribution is a simple product of $N+1$ terms, each of which corresponds to a single sequential transition along the ring. In the time domain, one has a corresponding convolution of exponentials 
\be
f_i(t)\equiv \Gamma_ie^{-\Gamma_i t}
\ee
which each on its own describe the elementary process of independent random transitions. 
The fact that $\hat{w}(z\to\infty) \sim \Gamma_0 ...\Gamma_N z^{-(N+1)}$ leads to the short-time expansion
\be\label{chainshortwt}
w(\tau) \sim \frac{\Gamma_0...\Gamma_N}{N!}\tau^N,\quad \tau\to 0, 
\ee
which reflects the $N+1$ elementary transitions with probabilities $\Gamma_i$ within the system, cf. \eq{shorttimeform}.

\subsubsection{Single Level Quantum Dot}
The case $N=1$    describes the single resonant level model in the large bias limit \cite{BB00}. For example the waiting time distribution $\hat{w}_{R}(z)$ of a single level quantum dot coupled to a left emitter and a right collector reservoir with corresponding tunnel rates $\Gamma_0=\Gamma_L$, $\Gamma_1=\Gamma_R$ and counting of electrons in the collector only is given by
\be
\label{wz_single}
 \hat{w}_{R}(z) = \frac{\Gamma_R}{z+\Gamma_R}\frac{\Gamma_L}{z+\Gamma_L}.
\ee
This expression has two poles at $z=-\Gamma_R$ and $z=-\Gamma_L$ and thus, in its pole structure, contains separate information on the two tunnel barriers (left and right). 
In contrast, the corresponding noise spectrum,
\be
S_{R}(\omega) = I \left( 1- \frac{2\Gamma_L\Gamma_R}{\Gamma^2+\omega^2}\right),\quad  I=\frac{\Gamma_L\Gamma_R}{\Gamma}
\ee
has two poles at $\omega =\pm i\Gamma$ which only depend  on the sum of both tunnel rates and not the individual tunnel rates.
Thus, already this simple example reveals that the physical information contained in the two quantities, $S_{R}(\omega)$ and $\hat{w}_{R}(z)$, is not the same: in order to extract both tunnel rates from the noise spectrum, one needs additional information such as the absolute value of the current $I$.

In the time domain, by Laplace back-transforming \eq{wz_single} one finds the result first obtained by Davies {\em et al.} \cite{DHHW92},
\be
w_{R}(\tau) =\Gamma_R\Gamma_L \frac{e^{-\Gamma_L\tau}-e^{-\Gamma_R\tau} }{\Gamma_R-\Gamma_L },
\ee
which has a short-time expansion $w_{R}(\tau) =\Gamma_R\Gamma_L\tau + O(\tau^2)$. The vanishing of $w_{R}(\tau)$ for $\tau=0$ indicates that after tunneling of an electron into the collector the dot is in the empty state an no other electron can follow immediately. The corresponding relaxation current \eq{relaxdef}
\be
I^{\rm relax}(t) = I \left[ 1- e^{-(\Gamma_R+\Gamma_L)t}\right]
\ee
describes the increase of the current from zero at $t=0$ (empty dot) towards the stationary current $I$.

\subsubsection{Large Ring $N\to \infty$}
Another interesting case is the ring, \eq{ring}, with identical  rates that are scaled up with increasing $N$ according to 
\be
\Gamma_i= (N+1) \gamma.
\ee
The waiting time distributions 
\be
\hat{w}_N(z)\equiv \left(1+\frac{z}{(N+1)\gamma}\right)^{-N}
\ee
then yield the same stationary average current $I=1/\langle \tau \rangle$ as 
\be
I = I_N\equiv 1/\hat{w}'_N(0)=\gamma
\ee
for each $N$, but the Full Counting Statistics following from \eq{FCS_single}, 
\be
z_0(\chi) = (N+1) \gamma\left(e^{\frac{i \chi}{N+1}}-1     \right), 
\ee
indicates that all the second and higher cumulants vanish for $N\to \infty$. 
The limit of $N\to\infty$ in fact  leads to {\em deterministic transport} without fluctuations, i.e. a waiting time distribution
$
 \hat{w}_\infty(z)\equiv \lim_{N\to\infty} \hat{w}_N(z) = e^{-z/\gamma}.
$
In the time-domain, the waiting times converge towards  a delta peak at the inverse transition rate $\gamma$,
${w}_\infty(\tau) = \delta\left(\tau-\frac{1}{\gamma}\right)$, and the `relaxation' current, i.e. the current for empty initial condition at time $t=0$, becomes a series of delta-peaks,
\be
{I}_{\infty}^{\rm relax}(t)&=& \sum_{n=1}^\infty \delta\left(t-n\frac{1}{\gamma}\right),
\ee
cf. \eq{Irelaxexpansion}, {where the term `relaxation' is of course misleading in this limit}. The deterministic character of the transport is confirmed by noting that the noise spectrum $S(\omega)$, \eq{novotny_formula_w}, is identical zero since
$1 + \sum_\pm   \left[\hat{w}_\infty(\pm i\omega)^{-1} -1 \right]^{-1}  =0$.

\subsection{Multi-Level Single Dot}\label{multileveldot}
The next example is  a single quantum dot with $N$ levels coupled to a left emitter and a right collector via tunnel rates
$\gamma_i$ (left) and $\Gamma_i$ (right),
\be
\mathcal{L} &=&\left(
  \begin{array}{ccccc}
-\gamma_L & \Gamma_1 & ... &  \Gamma_{N+1} &  \Gamma_N \\ 
 \gamma_1 & -\Gamma_1 & 0 &... & 0\\
 \gamma_2        & 0 & -\Gamma_2 &...& 0\\
 ... &  ...  & ... & ... & ...\\  
 \gamma_N  & 0 & ...       & 0 & -\Gamma_N\\
\end{array}\right)\nonumber\\
\gamma_L&\equiv&\gamma_1+...+\gamma_N.
\ee
In the strong Coulomb blockade regime, single electrons occupy one of the $N$ levels at a time, and transport is  from the left to the right with an infinite bias between emitter and collector. 
This is a single-reset system with jump operators $\mathcal{J}_\alpha=\ket{\alpha}\bra{\alpha}$, $\alpha=L/R$ (left/right), where $\bra{R}\equiv(0,\Gamma_1,\Gamma_2,...,\Gamma_N)$,  $\ket{R}=(1,0,...,0)^T$, and 
$\ket{L}\equiv \gamma_L^{-1}(0,\gamma_1,\gamma_2,...,\gamma_N)$,  $\bra{L}=(\gamma_L,0,...,0)^T$. Note that
$\bra{R} R\rangle\rangle = \bra{L} L\rangle\rangle =0$ and therefore $\mathcal{J}_R^2=\mathcal{J}_L^2=0$. 
The waiting time distribution for counting  only on the left or only on the right side, $\hat{w}_L(z)=\hat{w}_R(z)\equiv \hat{w}(z)$, is thus
\be
\label{wz_Nlevel}
 \hat{w}(z) &=& \bra{1} (z-\mathcal{L}_0)^{-1}\ket{1} = \sum_{i=1}^N \Gamma_i \left[(z-\mathcal{L}_0)^{-1}\right]_{i0}\nonumber\\
            &=& \sum_{i=1}^N \frac{\gamma_i\Gamma_i}{(z+\gamma_L)(z+\Gamma_i)},
\ee
{which again follows from considering the ratio of determinants}. 
In contrast, a splitting $\mathcal{L}=\mathcal{L}_0+\mathcal{J}_L+\mathcal{J}_R$ for counting on both sides leads to $\hat{w}_{LL}(z)=\hat{w}_{RR}(z)=0$ owing to the single reset character of the system: a tunneling event on the right side can only be followed by a tunneling event on the left side, and vice versa.

The sum over all levels $i$ in \eq{wz_Nlevel} reflects the `parallel' character of transport through the multi-level system: each level contributes with a single waiting time distribution (analogous to \eq{wz_single} but with the left tunnel rate $\Gamma_L$ replaced by the sum $\gamma_L$ of all left tunnel rates $\gamma_i$). For example, if all rates are identical, $\gamma_i=\Gamma_i=\gamma$, one is back to an effective single-level case, \eq{wz_single} with $\Gamma_R=\gamma$ and $\Gamma_L=N\gamma$ which for large $N$ leads to purely Poissonian noise properties.

Some other interesting features of this system can be extracted from \eq{wz_Nlevel}. The case $N=2$ with strongly asymmetric rates  leads to dynamical channel blockade as characterized by huge Fano factors and super-Poissonian noise \cite{KWS03,Bel05,LAB07}:  the transport is fast on the time-scale of the inverse of the smaller rate, but the electron is occasionally trapped in the `slow' level.
The Fano factor $F$ is easily obtained via the second moment of the waiting times \cite{DHHW92}, $F = \langle \tau^2 \rangle/ \langle \tau \rangle^2 -1$, and
for identical left and right rates $\gamma_i=\Gamma_i$ one has
\be
F =  \frac{\hat{w}''(0)}{\hat{w}'(0)^2 }-1 = \frac{1}{9}\left( 1 + 2 \frac{\gamma_1}{\gamma_2} + 2 \frac{\gamma_2}{\gamma_1} \right),
\ee
which increases with increasing asymmetry between the two rates.

\subsection{Double Quantum Dot (Strong Coulomb Blockade)}

\begin{figure}[t]
\centerline{\includegraphics[width=0.5\columnwidth]{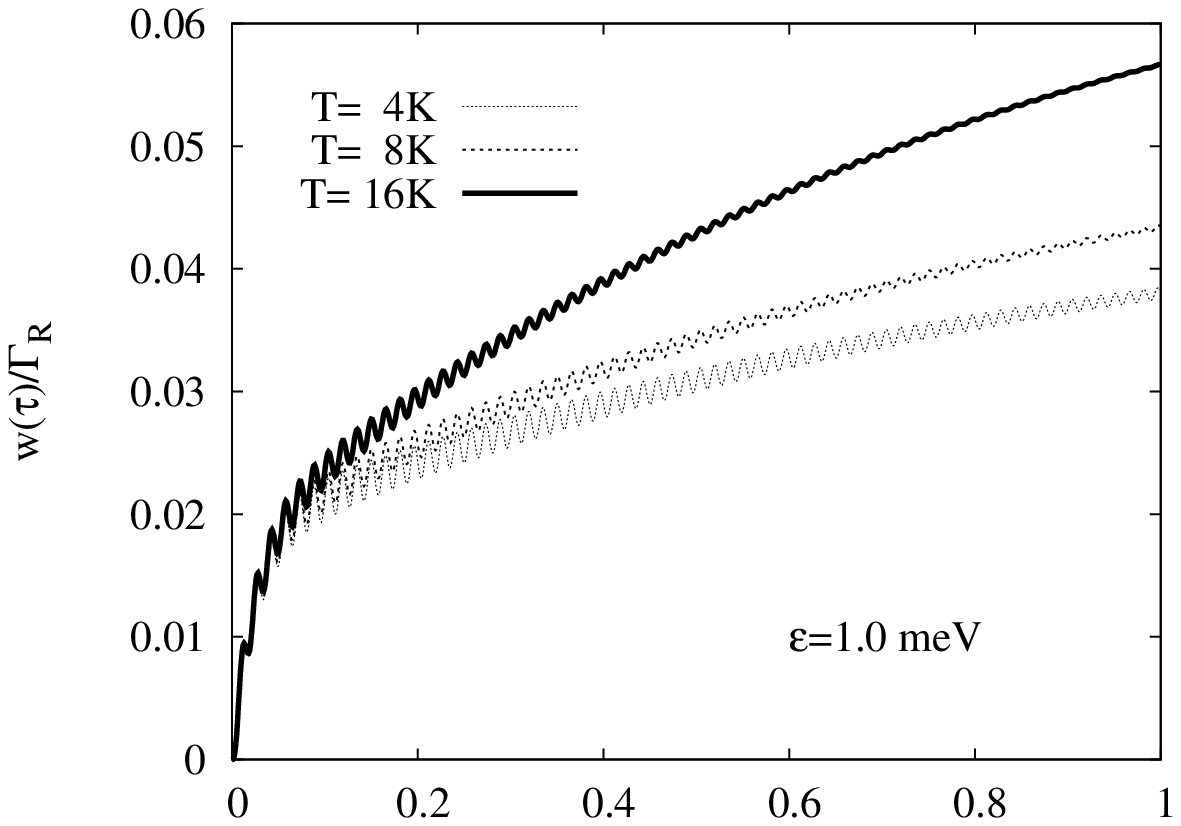}\includegraphics[width=0.5\columnwidth]{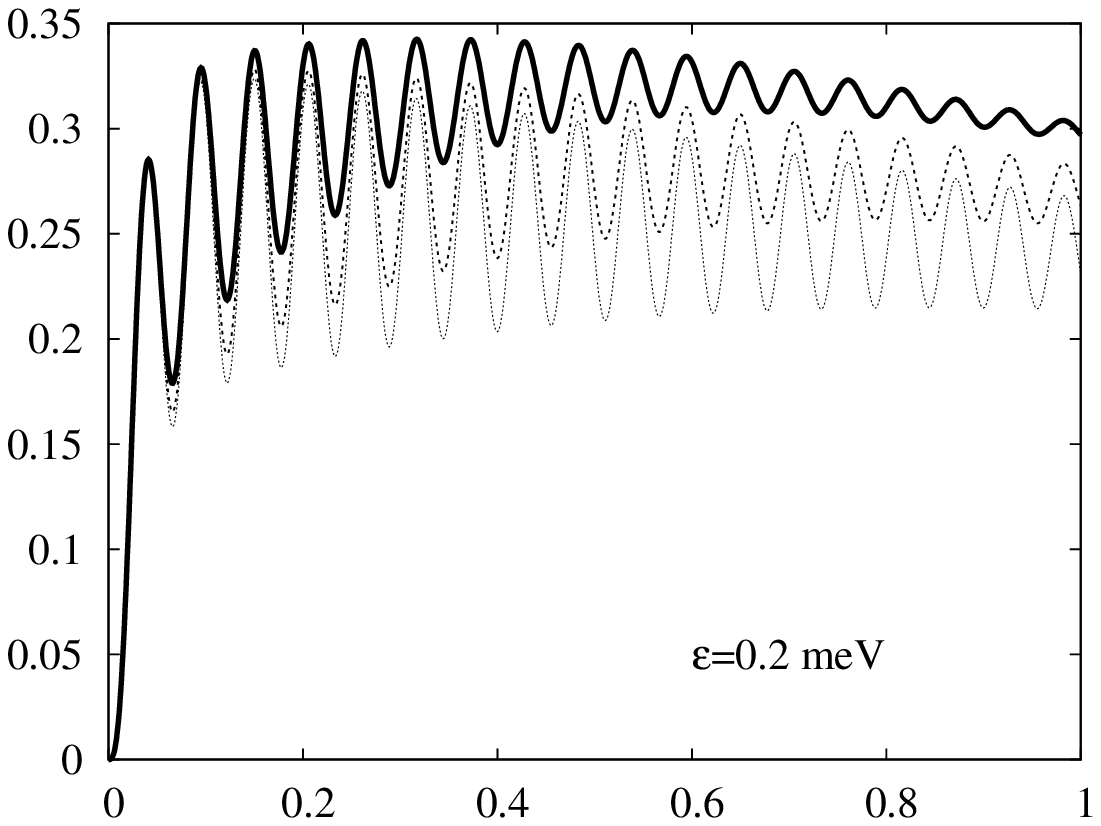}}
\centerline{\includegraphics[width=0.5\columnwidth]{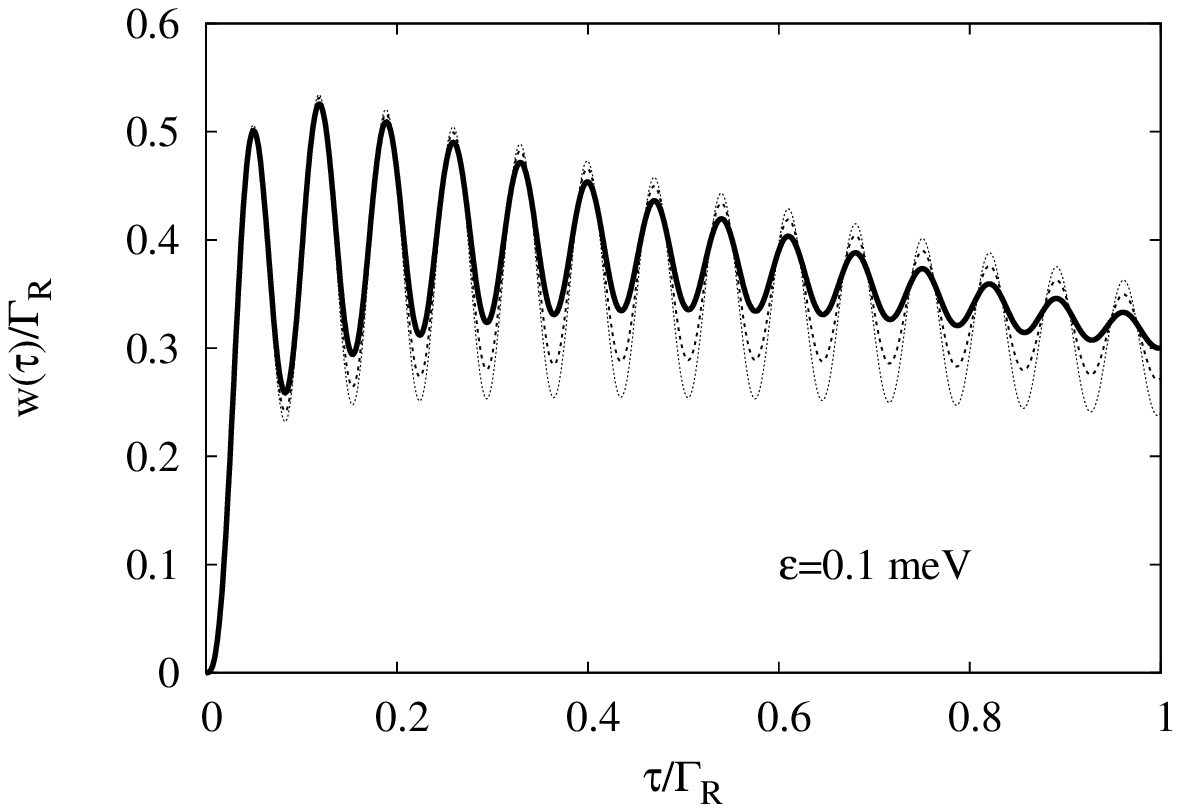}\includegraphics[width=0.5\columnwidth]{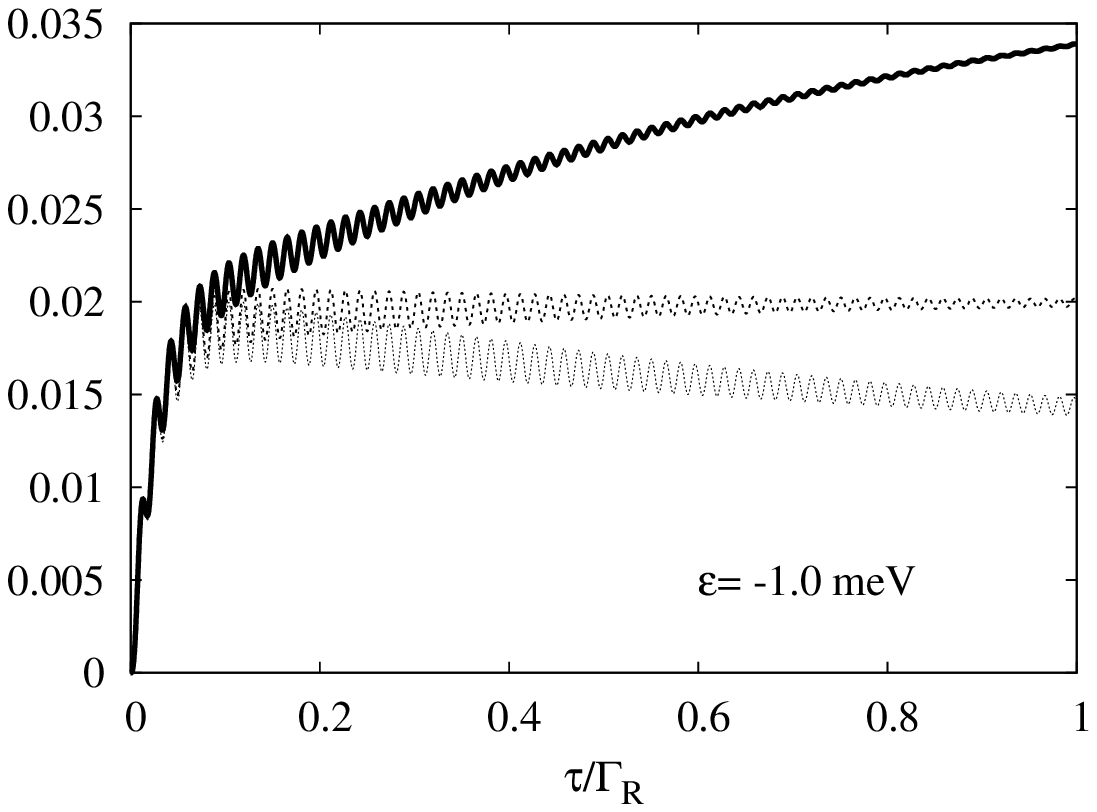}}

\caption[]{\label{double} Waiting time distribution $w(\tau)$ for transport through double quantum dot at temperatures $T=4,8,16$ K at $\varepsilon=1$meV, $0.2$meV, $0.1$meV, $-1.0$meV (clockwise). Right tunnel rate $\hbar\Gamma_R=2.5\mu$eV, other parameters $\hbar\Gamma_L=0.1$meV, $T_c=0.1$meV, electron-phonon coupling parameter $g=0.002$, cutoff $\hbar\omega_c=5$meV.}
\end{figure} 
This is the simplest model where quantum coherence becomes visible in transport \cite{SN96,GP96,Gur98,BK99,Bra05,KSWS06,Kieetal07}. Spin-polarized electrons move between two tunnel-coupled levels $|L\rangle$ (left) and $|R\rangle$ (right) attached to fermionic reservoirs. The Hamiltonian is a transport version of the spin-boson model ($\hbar=1$), 
\begin{eqnarray}\label{model}
{\cal H}&=&{\cal H}_{S}+{{\cal H}_{\rm res}}+{\cal H}_{T}+{\cal H}_{ep}  + {\cal H}_{p}    \\
{\cal H}_{S}&=&\frac{\varepsilon}{2} {\hat{\sigma}_z} + {T_c}\hat{\sigma}_x ,\quad {\cal H}_{\rm res}=\sum_{k,\alpha=L,R}\varepsilon_k c_{k,\alpha{}}^{\dagger}c_{k,\alpha{}}^{\phantom{\dagger}} \nonumber\\
{\cal H}_T&=&\sum_{k,\alpha=L,R} (V_k^\alpha{} c_{k,\alpha{}}^{\dagger}|0 \rangle \langle \alpha|+H.c.) \nonumber \\
{\cal H}_{ep}&=&\hat{\sigma}_z \sum_Q \frac{g_Q}{2}\left( a_{-Q} + a_Q^{\dagger}     \right),\quad 
{\cal H}_{p} =  \sum_Q  \omega_Qa_Q^{\dagger}a_{Q},  \nonumber
\end{eqnarray}
with pseudo-spin $\hat{\sigma}_z\equiv |L\rangle \langle L| - |R\rangle \langle R|$, $\hat{\sigma}_x\equiv |L\rangle \langle R| + |R\rangle \langle L|$,
the `empty' state $|0\rangle$, the standard tunnel Hamiltonian ${\cal H}_T$ for coupling to the reservoirs ${\cal H}_{\rm res}$, and coupling of the transport electron in the double dot to a phonon bath ${\cal H}_{p}$ via ${\cal H}_{ep}$. 
One can derive a generalized Master equation in the limit of infinite source-drain bias \cite{GP96,Gur98,SN96} and in the regime of strong Coulomb blockade, i.e. with only one additional transport electron in the double dot. 

The Liouvillian in the basis $\rho= (\rho_0,\rho_L,\rho_R,\Re \rho_{RL}, \Im  \rho_{RL})$
{has the form}
\begin{eqnarray}
\mathcal{L} &=&\left(
\begin{array}{ccccc}
-\Gamma_L & 0 & \Gamma_R  & 0 & 0\\
\Gamma_L & 0 & 0 & 0 & 2T_c\\
0 & 0 & -\Gamma_R & 0 & -2T_c \\
0 & \gamma_+ & -\gamma_- & -\frac{\Gamma_R}{2} -\gamma & -\varepsilon\\
0 & -T_c & T_c & \varepsilon & -\frac{\Gamma_R}{2} -\gamma
\end{array}
\right)\nonumber\\
\label{eq:master1}
\end{eqnarray}
with tunnel rates $\Gamma_\alpha=2\pi\sum_{k_\alpha} |V_k^\alpha|^2\delta(\varepsilon-\varepsilon_{k_\alpha})$, $\alpha=L/R$, assumed as energy-independent rates for  electron-phonon interaction\cite{BV02,Bra05,Kieetal07}
\begin{eqnarray}
\gamma &=&\frac{g\pi}{\Delta^2}\left[\frac{\varepsilon^2}{\beta}+2T_c^2\Delta e^{-\Delta /\omega_c}
\coth{\left(\frac{\beta\Delta}{2}\right)}\right]\\
\gamma_{\pm} &=&g\frac{\pi T_c}{\Delta^2}\left[\frac{\varepsilon}{\beta}
-\frac{\varepsilon}{2}\Delta e^{-\Delta /\omega_c}\coth{\left(\frac{\beta\Delta}{2}\right)
\mp\frac{\Delta^2}{2}e^{-\Delta /\omega_c}}\right]\nonumber
\end{eqnarray}
with a dimensionless coupling constant $g$, a Debye cutoff $\omega_c$, the level splitting $\Delta = \sqrt{\varepsilon^2+4T_c^2}$, and the inverse temperature $\beta = (k_BT)^{-1}$. These electron-phonon rates correspond to a bosonic environment with Ohmic spectral density $\rho (\omega )=g\omega e^{-\omega /\omega_c}\Theta (\omega )$.

This is a single-reset system with a jump operator describing the tunneling of single electrons from the right dot into the collector, $\mathcal{J}_R\equiv\ket{R}\bra{R}$ with $\bra{R}\equiv (0,0,\Gamma_R,0,0)$ and $\ket{R}\equiv (1,0,0,0,0)^T$. The relaxation current $\hat{I}^{\rm relax}(z)$ and thus, via \eq{Iwrelation}, the waiting time distribution $\hat{w}(z)$, is given by \cite{Bra05}
\begin{widetext}
\be
\hat{I}^{\rm relax}(z)&\equiv &\Gamma_R\hat{n}_R(z)
\equiv \frac{\Gamma_R\Gamma_Lg_+(z)}{z\left\{
[z+\Gamma_R+g_-(z)](z+\Gamma_L)+(z+\Gamma_R+\Gamma_L)g_+(z)\right\}}\\
g_{\pm}(z)&\equiv&
  2T_c\frac{T_c(\gamma+ \Gamma_R/2+z)-\varepsilon\gamma_{\pm}}{(\gamma+ \Gamma_R/2+z)^2+\varepsilon  ^2}.
\ee
\end{widetext}
The stationary current $I$ follows directly from $\hat{I}^{\rm relax}(z\to 0) \sim I/z$, or via the stationary state $\rho_0$ that is either obtained from the eigenvector of $\mathcal{L}$ with eigenvalue zero, or  from solving the linear equation $\mathcal{L}'\rho_0=(1,0,0,0,0)^T$, where $\mathcal{L}'$ is obtained by replacing the first row of the singular matrix $\mathcal{L}$ with $(1,1,1,0,0)$, corresponding to the normalization condition $n_0+n_L+n_R=1$.

The five poles of $\hat{w}(z)$ are obtained analytically via
\be
\hat{w}(z) = \frac{\hat{I}^{\rm relax}(z)}{1+\hat{I}^{\rm relax}(z)}
\ee
and in the case of no electron-phonon scattering are {explicitly given by}
\be
\label{five_roots}
z_1&=&  -\Gamma_L \nonumber\\
z_{2/3} &=& -\frac{\Gamma_R}{2}\pm \frac{\Delta_R}{\sqrt{2}} \sqrt{\sqrt{1+\left(\frac{\Gamma_R\varepsilon}{\Delta_R}\right)^2}-1}\nonumber\\
z_{4/5} &=& -\frac{\Gamma_R}{2}\pm i\frac{\Delta_R}{\sqrt{2}} \sqrt{\sqrt{1+\left(\frac{\Gamma_R\varepsilon}{\Delta_R}\right)^2}+1}\nonumber\\
\Delta_R &\equiv& \sqrt{\varepsilon^2+4T_c^2-\frac{\Gamma_R^2}{4}}.
\ee
One recognizes that the information on the left tunnel barrier contained in the pole $z_1=  -\Gamma_L$ is completely separated from the other poles. This  is a consequence of the useful relation \eq{twosplit} for an unraveling according to counting electrons both in the left and right leads,
\be
\hat{{w}}(z) &=& \hat{w}_{RR}(z) + \frac{\hat{w}_{RL}(z)\hat{w}_{LR}(z)}{1-\hat{w}_{LL}}\nonumber\\
&=& \hat{w}_{RL}(z)\hat{w}_{LR}(z)= \hat{w}_{RL}(z) \frac{\Gamma_L}{z+\Gamma_L},
\ee
because two subsequent jumps in the emitter (L) or collector (R) are not possible due to the strong Coulomb blockade assumption, i.e. $\hat{w}_{LL}=\hat{w}_{RR}=0$. A jump on the right side leaves the system empty, with the trivial re-charging process as described by $\hat{w}_{LR}(z)=\frac{\Gamma_L}{z+\Gamma_L}$ following, and thus all the relevant information on the quantum system is contained in $\hat{w}_{RL}(z)$ which describes the dynamics after a jump into the double dot from the left lead.  

In the time domain, it is easier to obtain $w(\tau)$ directly from its definition \eq{wdefinition}. The result shown in Fig. \ref{double} was produced using the matrix exponential in MATHEMATICA. Parameters were chosen close to those used in a recent experimental and theoretical analysis of the Fano factor in vertical double quantum dots \cite{Kieetal07}. The most important feature in $w(\tau)$ is the appearance of oscillations with period $\approx \Delta$ for $\Delta\gg \Gamma_R$, which are due to the coherent coupling $\propto T_c$ between the two quantum dots as reflected in the two imaginary parts in the zeroes \eq{five_roots}. With increasing temperature $T$ of the phonon bath, the oscillations become less pronounced although they are still visible at relatively large $T$.


The short time expansion of $w(\tau)$ is obtained via $\hat{w}(z\to\infty) \sim 2 T_c^2\Gamma_R\Gamma_L z^{-4}$ or directly from \eq{shorttimedirect},
\be
w(\tau) = \frac{1}{3!}2T_c^2\Gamma_R\Gamma_L  \tau^3 + O(\tau^4),
\ee
which reflects the elementary transitions within the system: tunneling of an electron from the left lead to the left dot $\propto \Gamma_L$ followed by  coherent tunneling from the left dot to the right dot $\propto T_c^2$, and finally tunneling of an electron from the right dot to the right lead $\propto \Gamma_R$. 
 The cubic behaviour, $w(\tau)\propto \tau^3$, for coherent tunneling is in contrast to {\em sequential} tunneling through $N=2$ dots (ring example above) where one would find a quadratic behaviour of $w(\tau)$ at small $\tau$, cf. \eq{chainshortwt}.

\subsection{Example: Three-State System}
As  a non-trivial multiple-reset system,  consider the Anderson single-impurity model in the limit of infinite bias,
\be
\label{Anderson}
\dot{p}_0&=& -\gamma_{L\uparrow} p_0 + \gamma_{R\uparrow }  p_\uparrow +  \gamma_{R\downarrow }  p_\downarrow \nonumber\\
\dot{p}_\uparrow &=& \gamma_{L\uparrow} p_0 -(\gamma_{R\uparrow}+\Gamma_{L\uparrow} )  p_\uparrow +  \Gamma_{R\uparrow}  p_2 \nonumber\\
\dot{p}_\downarrow &=& \gamma_{L\downarrow} p_0 -(\gamma_{R\downarrow}+\Gamma_{L\downarrow} )  p_\downarrow +  \Gamma_{R\downarrow}  p_2 \nonumber\\
\dot{p}_2 &=& - \Gamma_{L\uparrow} p_\uparrow - \Gamma_{L\downarrow}  p_\downarrow - (\Gamma_{R\uparrow}+\Gamma_{R\downarrow} )   p_2 .
\ee
The system consist of a single electronic level with four electronic states: empty ($0$), spin up/down ($\uparrow$,$\downarrow$) and doubly occupied ($2$), and is again coupled to an emitter (left) and a collector (right). In the infinite voltage limit, rates $\gamma_{\alpha\sigma}$ describe transitions between empty and singly occupied states with spin $\sigma$, and  rates $\Gamma_{\alpha\sigma}$ describe transitions between singly and doubly occupied states. Assuming spin-independent rates, one can introduce $p_1\equiv p_\uparrow + p_\downarrow$ and thus in the components 
$\rho=(p_0,p_1,p_2)$, the Liouvillian is
\be\label{three_Liouville}
\mathcal{L} 
&=&
\left(
\begin{array}{ccc}
 -2\gamma_L &    e^{i\chi_R} \gamma_R  & 0\\
2\gamma_L e^{i\chi_L} & -(\Gamma_L +\gamma_R) &    e^{i\chi_R} 2\Gamma_R     \\
 0       & \Gamma_L  e^{i\chi_L} & -2\Gamma_R  \\ 
\end{array}
\right)_{\chi_L =\chi_R=0}
\ee
where the counting fields $e^{i\chi_{R/L}}$  indicate the jump operators $\mathcal{J}_{R/L}$. 
For example, $\mathcal{J}_{R}=\ket{R1}\bra{R1} +\ket{R2}\bra{R2}$ with kets $\ket{R1}=(1,0,0)^T$, $\ket{R2}=(0,1,0)^T$,  and bras $\bra{R1}=(0,\gamma_R,0)$, $\bra{R2}=(0,2\Gamma_R,0)$.
This is thus a multiple (twofold) reset system with
$\mathcal{J}_{R/L}^2\ne 0$ and $\mathcal{J}_{R/L}^3= 0$, which reflects the two possible reset states (singly occupied and empty) contributing to transport. One can check by direct calculation that the FCS equation, \eq{FCS_multiple}, as obtained from the two by two waiting time matrix $\mathbf{W}(z)$ (e.g., for counting in the collector) coincides with the  usual eigenvalue equation, $\det [\mathcal{L}(e^{i\chi_R}) -z]=0 $.
\begin{figure}[t]
\includegraphics[width=\columnwidth]{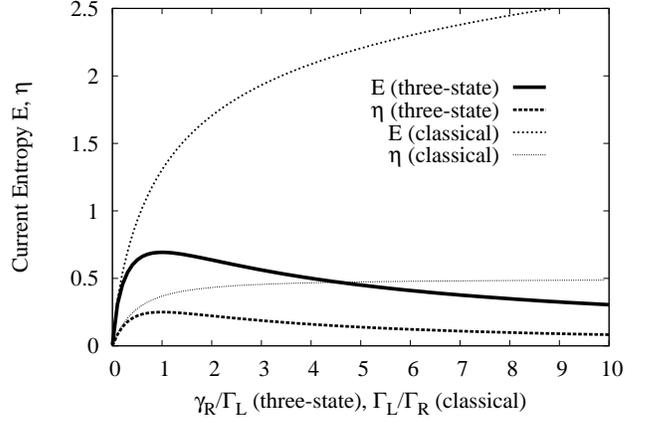}
\caption[]{\label{entropy}
Comparison between current entropy $E$  and waiting time parameter $\eta$ as a function of  $\gamma_R/\Gamma_L$ [$\Gamma_L/\Gamma_R$]    for the three-state system \eq{Anderson} [the classical model \eq{classical} ].
}
\end{figure} 

The stationary state is obtained as
\be
p_1 &=& 2\frac{\gamma_L}{\gamma_R} p_0,\quad p_2 = \frac{\gamma_L}{\gamma_R}\frac{\Gamma_L}{\Gamma_R} p_0\nonumber\\
p_0 &=& \frac{1}{1+ 2 \frac{\gamma_L}{\gamma_R} + \frac{\gamma_L}{\gamma_R}\frac{\Gamma_L}{\Gamma_R}},
\ee
with a stationary current 
$I = 2\gamma_L(1+ \frac{\Gamma_L}{\gamma_R})p_0$. 
The (reduced) waiting time distribution for measuring on either the left or the right side only, \eq{wreducedef},  follows as 
\be
&  &\hat{w}^{(r)}(z)\equiv \hat{w}_L^{(r)}(z)=  \hat{w}_R^{(r)}(z)  \\
&= &\frac{\left(z \Gamma _L+2 \gamma _L \left(\gamma
   _R+\Gamma _L\right)\right) \left(2 \Gamma _L \Gamma
   _R+\gamma _R \left(z+2 \Gamma
   _R\right)\right)}{\left(z+2 \gamma _L\right)
   \left(\gamma _R+\Gamma _L\right) \left(z+\gamma
   _R+\Gamma _L\right) \left(z+2 \Gamma _R\right)}.\nonumber
\ee
The short-time limit, \eq{shorttimeform}, therefore is 
\be
w^{(r)}(\tau\to 0) = \frac{\Gamma_L}{1+\Gamma_L/\gamma_R} + O(\tau),
\ee
which indicates that there is a finite probability to observe two electrons just one after another, in contrast to single-reset systems.

On the other hand, the waiting times for counting in both leads, corresponding to a splitting $\mathcal{L}=\mathcal{L}_0 + \mathcal{J}_L +\mathcal{J}_R$, lead to 
\be
\label{eta_three}
\eta\equiv \hat{w}_{LL}(0)=\hat{w}_{RR}(0) = \frac{\gamma_R/\Gamma_L}{(1+   \gamma_R/\Gamma_L)^2  }.
\ee
In contrast to the single-reset systems discussed above, this means that there can be two subsequent electron jumps on either side. The quantity $\eta>0$ offers itself as an experimentally accessible way to quantify the multiple-reset character of the system. In fact, $\eta$  can be compared to the current entropy, \eq{current_entropy},  introduced above,
\be
\label{entropy_three}
E\equiv E_R=E_L = \frac{\log (1+\gamma_R/\Gamma_L )}{1+   \gamma_R/\Gamma_L} +(L\leftrightarrow R).
\ee
It is interesting to notice that both quantities, $\eta$ and $E$, depend on the same combination of rates and in fact display a qualitatively quite similar behaviour, cf. Fig.\ref{entropy}.

\begin{figure}[t]
\includegraphics[width=\columnwidth]{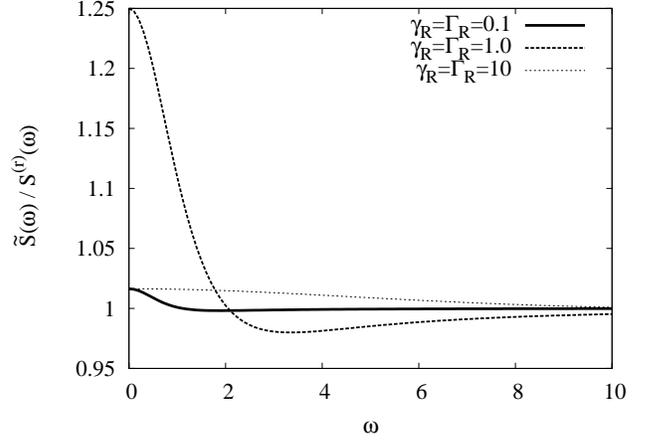}
\caption[]{\label{fidelity} Ratio of the spectrum  $\tilde{S}(\omega)$, \eq{F0def}, and the proper (reduced)  noise spectrum $S^{(r)}(\omega)$, \eq{Sreduced},  for the three-state system \eq{Anderson}. Frequency $\omega$ and tunnel rates are in units of $\Gamma_L=\gamma_L$.}
\end{figure}

A further way to quantify the multiple reset character of transport is to exploit the fact that the reduced noise spectrum, $S^{(r)}(\omega)$, is {\em not} simply obtained via
the {reduced} waiting time distribution $\hat{w}^{(r)}(z)$. One can therefore introduce a `fidelity'
\be\label{F0def}
F(\omega)&\equiv& \frac{\tilde{S}(\omega)}{S^{(r)}(\omega)}\\
\tilde{S}(\omega) &=& I\left[  1 + \sum_\pm   \left[\hat{w}^{(r)}(\pm i\omega)^{-1} -1 \right]^{-1} \right],
\ee
which is unity for single reset systems and tends towards one for multiple-reset system at large frequencies $\omega$ as a consequence of \eq{highfreduced}. The behaviour of $F(\omega)$ at various right tunnel rates $\gamma_R=\Gamma_R$ for fixed $\gamma_L=\Gamma_L$ is shown in Fig. \ref{fidelity}. The deviation of $F(\omega)$ from unity is strongest for $\gamma_R/\Gamma_L\approx 1$. This corresponds well with the current entropy maximum in Fig. \ref{entropy}, although the latter only depends on the ratio $\gamma_R/\Gamma_L$ whereas $F(\omega)$ depends on all four tunnel rates independently. 

The limit of very large and very small $\gamma_R$ can be understood from the waiting time distribution, i.e. 
$\lim_{\gamma_R\to \infty} \hat{w}(z)= \frac{2\Gamma_L}{2\Gamma_L+z}$,
which corresponds to the empty stationary state with Poissonian waiting time distribution, and 
$\lim_{\gamma_R\to 0} \hat{w}(z)= \frac{2\Gamma_L\Gamma_R}{(2\Gamma_R+z)(\Gamma_L+z)}$,
which corresponds to a stationary state $\rho_0=(0,2\Gamma_R/\Gamma,\Gamma_L/\Gamma)$, $\Gamma=\Gamma_L+\Gamma_R$ of a two-state system (one or two electrons). In both these limits, $F(\omega)$ tends towards unity in agreement with Fig. \ref{fidelity}.

\subsection{Classical Transport}
The final example is a system which can accommodate an arbitrary large number $n$ of particles, $n=0,1,2...$ with occupation probability $p_n$ in the stationary state $\rho=(p_0,p_1,p_2,p_3,...)$. Additional single particles enter the system at the rate $\Gamma_L$ from a left reservoir regardless of $n$, and single particles leave the system into the right reservoir at rate $\Gamma_R$, leading to system transitions $n\to n-1$ at a rate $n\Gamma_R$. The particles are assumed classical, i.e. there are no effects due to boson or fermion statistics. The Liouvillian is
\be \label{classical}
\mathcal{L} =\left(
  \begin{array}{llllll}
-\Gamma_L & \Gamma_R & 0 &... & 0 & 0 \\ 
 \Gamma_L & -\Gamma_R-\Gamma_L & 2\Gamma_R & & ... & 0\\
 0        & \Gamma_L & -2\Gamma_R-\Gamma_L & 3\Gamma_R& ... &0\\
 ... &  ...  & ... & ... & ...&...\\  
\end{array}\right),
\ee
(infinitely many entries). The average number $\bar{n}_t$ of particles in the system as a function of time $t$ simply obeys
\be
\frac{d}{dt}\bar{n}_t = \Gamma_L -\Gamma_R \bar{n}_t.
\ee
One decomposes $\mathcal{L}=\mathcal{L}_0+ \mathcal{J}_R+ \mathcal{J}_L$ with jump operators with matrix elements 
\be
\left[\mathcal{J}_R\right]_{kl} &=& k \Gamma_R \delta_{k,l-1},\quad \left[\mathcal{J}_L\right]_{kl} = \Gamma_L \delta_{k,l+1}.
\ee
The system has a stationary state $\rho_0$ given by the Poisson distribution
\be
\label{statclassic}
p_n = \frac{\alpha^n}{n!}e^{-\alpha},\quad \alpha \equiv \frac{\Gamma_L}{\Gamma_R}.
\ee
Using the stationary current  $I=I_R=I_L=\Gamma_L$ and the stationary state $\rho_0$ with \eq{statclassic}, 
the waiting time distributions are obtained by simple matrix multiplication in  $\hat{w}_{ij}(z)= { \mbox{\rm Tr} \mathcal{J}_i\mathcal{W}_j(z)\rho_0}/I$,
\be
\label{classic_gamma}
\hat{w}_{LL}(z) = \hat{w}_{RR}(z) &=& \sum_{n=0}^{\infty} \frac{\alpha p_n}{{z}/{\Gamma_R}+\alpha+n+1}\nonumber\\
\hat{w}_{LR}(z) &=& \sum_{n=0}^{\infty} \frac{n p_n}{{z}/{\Gamma_R}+\alpha+n-1}\nonumber\\
\hat{w}_{RL}(z) &=& \sum_{n=0}^{\infty} \frac{(n+1) p_n}{{z}/{\Gamma_R}+\alpha+n+1},
\ee
which can be expressed in terms of (incomplete) Gamma functions and which correspond to a measurement on both the right and the left side. A measurement on, e.g., the left side only corresponds to a splitting  $\mathcal{L}=\mathcal{L}_0 +\mathcal{J}_L$ with a single jump operator $\mathcal{J}_L$ only. The corresponding waiting time distribution, $\hat{{w}}_L(z)$, {has to fulfill} the identity \eq{twosplit} between the two splittings from Sect. \ref{sectionsplitting}, 
\be
\hat{{w}}_L(z) = \hat{w}_{LL}(z) + \frac{\hat{w}_{LR}(z)\hat{w}_{RL}(z)}{1-\hat{w}_{RR}(z)},
\ee
which by explicit calculation using \eq{classic_gamma} yields
\be
\hat{{w}}_L(z) = \frac{\Gamma_L}{z+\Gamma_L},
\ee
which indeed is the expected result as the particles on the left side enter the system independently. Since $\hat{w}_{LL}(z) =\hat{w}_{RR}(z) $, one obtains the same result,  
$\hat{{w}}_L(z)=\hat{{w}}_R(z)$, for counting on the right side. 

There are an infinite number of reset states with corresponding currents 
$I_{kL}\equiv p_k \Gamma_L$ for $k=0,1,2,...$ and $I_{kR} \equiv  p_k k \Gamma_R$ for $k=1,2,... $. Using $k p_k = \alpha p_{k-1}$, \eq{statclassic}, the current entropies, \eq{current_entropy}, are given by the entropy of the Poisson distribution $p_n$ itself,
\be
\label{entropy_classical}
E\equiv E_R = E_L = \mathcal{S}[p_n]= -\sum_n p_n \log p_n.
\ee
which is positive for  $\alpha\equiv \Gamma_L/\Gamma_R>0$. 
On the other hand, the quantity
\be
\label{eta_classical}
\eta \equiv \hat{w}_{LL}(0) = \hat{w}_{RR}(0)
\ee
is non-zero as there is a finite probability for yet another transition into or out of the system immediately following a previous one. Both the entropy 
$E$ and the zero-frequency waiting time $\eta$ thus reflect the multiple-reset character of the system. As a function of the asymmetry parameter $\alpha$, both 
are functions that increase from $0$ at $\alpha=0$ with growing $\alpha$, {cf. Fig. \ref{entropy}}, but  $\eta$ has an asymptote at $\frac{1}{2}$ whereas $E$ continues to grow logarithmically \cite{Evans} as $E\sim \frac{1}{2}\log(2\pi e\alpha)$.

\section{Remarks and Conclusion}
Other multiple-reset situations include systems with internal degrees of freedom. Transport of electrons through a single level with strong coupling to a vibrational degrees of freedom leads to a block structure of the Liouvillian: the simple (scalar) entries  in the two-by-two matrix of the $N=1$ case, \eq{ring}, become matrices, where the relatively complicated structure of the matrix elements leads to an avalanche-type of transport with  non-trivial power-laws in the noise spectrum $S(\omega)$ \cite{KRO05}. It would be interesting to find other systems where similar features can be extracted from waiting times of   Liouvillians with a relatively complex structure.

As a conclusion, the waiting times $w(\tau)$ appear to be a flexible theoretical tool for describing single particle transport, in particular as they contain the other statistical quantities - FCS and noise spectrum $S(\omega)$ - that have mainly been used in the past. Depending on the system, measuring  $w(\tau)$ could  provide additional information or a least serve as a cross-check for FCS and noise data. 

One interesting  generalization of the formalism in Sect. II would be to consider
Master equations of integro-differential type, e.g. containing non-Markovian memory kernels and retardation effects due to, e.g., strong electron-phonon coupling \cite{AB04,Flietal08}. A further question is the connection between un-symmetrized noise spectra \cite{AK00} and $w(\tau)$, and the relevance of multi-time waiting time distributions ($n>2$ jump operators in \eq{rho_cond}) and their relation to 
higher order frequency-dependent cumulants \cite{Emetal07}.

The author acknowledges T. Novotn\'{y} for useful hints concerning the formalism, interpretation and some references related to this manuscript. Discussions with R. Aguado, C. Flindt, C. Emary,  G. Kie{\ss}lich, D. Marcos, and G. Schaller are acknowledged. This work was supported by DFG grant BR 1528/5-1, the WE Heraeus foundation, and the DAAD.


\end{document}